\documentclass[prc,superscriptaddress,showpacs,amsmath,amssymb,twocolumn,preprintnumbers]{revtex4-1}
\usepackage{amsmath, amsthm, amssymb, amsfonts, braket,dcolumn}
\usepackage{hyperref}
\usepackage[dvipdfmx]{graphicx,color}

\begin{document}
\title{KIDS density functional for deformed nuclei: Examples of the even-even Nd isotopes}
\author{Hana Gil}
\address{Center for Extreme Nuclear Matter, Korea University, Seoul 02481, Korea}
\author{Nobuo Hinohara}
\address{Center for Computational Sciences, University of Tsukuba, Tsukuba 305-8577, Japan}
\address{Faculty of Pure and Applied Sciences, University of Tsukuba, Tsukuba 305-8571, Japan}
\author{Chang Ho Hyun}
\address{Department of Physics Education, Daegu University, Gyeonsan 38453, Korea}
\author{Kenichi Yoshida}
\email[E-mail me at: ]{kyoshida@ruby.scphys.kyoto-u.ac.jp}
\address{Department of Physics, Kyoto University, Kyoto 606-8502, Japan}

\begin{abstract}
\begin{description}
\item[Background]  A global description of the ground-state properties of nuclei in a wide mass range in a unified manner is 
desirable not only for understanding exotic nuclei but for providing nuclear data for applications. 

\item[Purpose]  We demonstrate the KIDS functional describes the ground states appropriately with respect 
to the existing data and predictions 
for a possible application of the functional to all the nuclei by taking Nd isotopes as examples. 

\item[Method] The Kohn--Sham--Bogoliubov equation is solved 
for the Nd isotopes with the neutron numbers ranging from 60 to 160 by employing the KIDS functionals 
constructed to satisfy both neutron-matter equation of state 
or neutron star observation and selected nuclear data. 

\item[Results] Considering the nuclear deformation improves the description of the binding energies and radii. 
We find that the discrepancy from the experimental data is more significant for neutron-rich/deficient isotopes and this can be made 
isotope independent by changing the slope parameter of the symmetry energy. 

\item[Conclusions] The KIDS functional is applied to the mid-shell nuclei for the first time. 
The onset and evolution of deformation are nicely described for the Nd isotopes. 
The KIDS functional is competent to a global fitting for a better description of nuclear properties in the nuclear chart.
\end{description}
\end{abstract}

\preprint{KUNS-2867}
\date{\today}

\maketitle

\section{Introduction}
Mass is a fundamental property of a particle. 
A systematic study of nuclear masses reveals 
novel features of atomic nuclei, and leads to a 
deeper understanding of the mechanism for the emergence of collective phenomena 
such as magic numbers, deformation, and superfluidity~\cite{lun03,bla06}. 
As the rare-isotope beam technology has developed, the playfield has been extended 
toward heavy and neutron-rich nuclei~\cite{nak17}. 
Exploring the limit of stability against the dissociation of nucleons is a frontier of the physics of exotic nuclei~\cite{nue19}. 
Although the proton drip line has been identified up to heavy elements~\cite{zha19}, 
the neutron drip line at $Z=9$ and $10$ was confirmed~\cite{ahn19} in nearly 20 years 
after the establishment at $Z=8$~\cite{tho13}. 
In certain years,  
most of the neutron-rich nuclei with larger $Z$ would remain unexplored experimentally, the progress notwithstanding. 
Therefore, it is desirable to develop a theoretical model applicable to nuclei in a vast region of the nuclear chart 
in a single framework.
Elucidating the properties of exotic nuclei further brings on 
an appreciation of the origin of heavy elements in the universe. 

There have been numerous attempts toward a global description of nuclear masses 
based on phenomenological models~\cite{kou05,liu11,ma15,ma19,he14}, 
the microscopic-macroscopic models~\cite{taj10,mol16,jia18,bha09,bha14}, the shell model~\cite{duf95}, 
the HFB model~\cite{gor09,gor16}, and using machine learning~\cite{niu18,niu19,she21}.
For a reliable description of unknown nuclei, 
there has been a rapid progress in modeling 
a microscopic approach considering a nuclear system 
as a many-body system interacting with nuclear forces.
An energy-density functional (EDF) method is such a model~\cite{taj96,sto03,erl12,afn13,zhe14,shi19,zha20}.

An EDF method is considered as the realization of density functional theory (DFT). 
A unified model of nuclear matter and finite nuclei, if it exists, 
would be constructed and developed in a similar way as DFT. At least
DFT does provide a reasonable basis for that. 
DFT is based on the Hohenberg--Kohn (HK) theorem~\cite{hoh64} which states 
that the ground state of a quantum many-body system 
is obtained by minimizing the EDF with respect to the density.
The HK theorem, however, does not provide the form of the EDF. 
An EDF for electronic systems has been constructed starting from the Coulomb force in a first-principle manner~\cite{yok21b}, 
while an {\it ab-inito} construction of 
the EDF for nuclear systems requires more effort~\cite{dru09,pug03,bha05,fur07,fur20,bra12,kem13,kem17a,lia18,yok19,yok19b,yok21}.
Therefore, one takes an alternative for the construction of the nuclear EDF by making full use of the existing 
measurement and observational data.

Most nuclear EDF widely adopted in the literature are constructed by 
building the functional from the expectation value of effective nuclear forces for the Slater determinant. 
There have been a few attempts to construct the nuclear EDF as a functional of the density directly 
and apply it to atomic nuclei~\cite{bcp08,bcpm13,meta17,bul18,PhysRevC.82.054307}, where 
the EDF is fitted to the equation-of-state (EoS) of nuclear matter and nuclear data. 

Controlling the uncertainty is an important issue in modeling. 
With an idea to make the  uncertainty quantification accessible in the nuclear structure theory, 
an expansion scheme in the low-energy few-nucleon effective field theory (EFT) 
has been employed in the construction of the EDF of nuclear matter.
This idea gave birth to the KIDS (Korea: IBS--Daegu--Sungkyunkwan) formalism.
The energy density is expanded in the power of $k_F/m_\rho$, with 
$k_F$ and $m_\rho$ being the Fermi momentum and the mass of $\rho$ meson, 
which is equivalent to the expansion in terms of the nucleon density to the one-third $\rho^{1/3}$.
Since this expansion is based on the perturbation theory, 
the KIDS formalism allows a quantitative estimation of the uncertainty and radius of the convergence.

The KIDS model was at first applied to homogeneous nuclear matter~\cite{pap18}.
In the subsequent applications, properties of closed-shell nuclei~\cite{gil19a, gil19b},
 and open-shell nuclei~\cite{gil20b} were calculated, where all the nuclei are assumed to be spherical.
The results are in good agreement with data on average. 
However, 
a non-negligible discrepancy between the calculation and experiment is observed 
in the neutron-rich unstable region where 
the deformation effect is suggested to appear experimentally~\cite{gil20a}.
In the present article, we investigate the deformation property with the KIDS functionals.
A recent work aiming at a global description of nuclear masses 
reported the calculation for the Nd isotopes in a relativistic EDF approach~\cite{zha20}.
The experimental data and the calculation in Ref.~\cite{zha20} 
provide a ground to test the performance of the KIDS model.
To this end, we solve the Kohn--Sham--Bogoliubov equation for the Nd isotopes 
with neutron numbers ranging from 60 to 160.
We see that the discrepancy found assuming the spherical symmetry is compensated dominantly by including deformation.
As a result, the agreement to experiment is improved, and it is as good as the result of the relativistic model~\cite{zha20}.

This paper is organized in the following way: 
the KIDS model is briefly recapitulated in Sec.~\ref{model};
the numerical details are also described.
Section~\ref{result} is devoted to the numerical results and discussion based on the model calculation; 
then, a summary is given in Sec.~\ref{summary}.

\section{Model}\label{model}

In the KIDS formalism, 
the model parameters are determined hierarchically so that
the parameters determined in later steps do not affect the physical properties considered in the prior steps.
In this work, the parameter-fitting steps are divided into the consideration of 
(i) nuclear matter,
(ii) closed-shell nuclei, 
and (iii) pairing.

In the first step, 
an energy per particle in homogeneous nuclear matter is expanded in the power of Fermi momentum as
\begin{align}
{\cal E}(\rho, \delta) = {\cal T}(\rho, \delta) + \sum_{i} (\alpha_i + \beta_i \delta^2 )\rho^{1+ i/3},
\label{eq:matter}
\end{align}
where ${\cal T}$ is the kinetic energy density, 
$\rho=\rho_{\text{n}}+\rho_{\text{p}}$ and $\delta = (\rho_{\text{n}} - \rho_{\text{p}})/\rho$.
Three $\alpha_i$ ($i=0,1$, and 2) and four $\beta_i$ $(i=0,1,2,$ and 3) fixed or fitted to selected nuclear matter properties are optimal for 
describing nuclear matter over a wide range of density~\cite{pap18,gil19b}.
In this work, we consider five KIDS models: KIDS0, KIDS-A, B, C, and D. 
For the KIDS0 model, three $\alpha_i$ are determined to produce the saturation density $\rho_0=0.16\, {\rm fm}^{-3}$,
the total energy per nucleon at the equilibrium $E_B = 16$ MeV,
and the nuclear matter incompressibility $K_0 = 240$ MeV,
 and four $\beta_i$ are determined to fit the pure neutron matter EoS in Ref.~\cite{akm98}.
The other four KIDS models, A, B, C, and D, are determined to satisfy both nuclear data and the recent neutron star observation~\cite{gil20a}.
The nuclear matter properties of the five KIDS models are summarized in Table~\ref{tab:model}.
For comparison with a standard Skyrme-force model, we include the results obtained with the SLy4 model~\cite{cha98}.
\begin{table}[t]
\caption{\label{tab:model} Parameters of nuclear matter EoS 
including the compression modulus $K_0$, and the symmetry energy parameters $J$, $L$, and $K_\tau$ are given in the units of MeV.}
\begin{ruledtabular}
\begin{tabular}{ldddd}
 & K_0 & J & L & K_\tau \\ \hline
KIDS0   & 240 & 32.8 & 49.1 & -375.1 \\
KIDS-A & 230 & 33 & 66 & -420 \\
KIDS-B & 240 & 32 & 58 & -420 \\
KIDS-C & 250 & 31 & 58 & -360 \\
KIDS-D & 260 & 30 & 47 & -360 \\
SLy4    & 229.9 & 32.0 & 45.9 & -322.8 
\end{tabular}
\end{ruledtabular}
\end{table}
%

\begin{table*}[t]
\caption{\label{tab:parameter} Skyrme force parameters of the KIDS models. 
We set $y_1 = y_2 = t_{33} = 0$.
The parameters are given in the units of MeV fm$^{3}$ for $t_0$, $y_0$, $t'_0$, 
MeV fm$^{4}$ for $t_{31}$, $y_{31}$,
MeV fm$^{5}$ for $t_1$, $t_2$, $t_{32}$, $y_{32}$,$W_0$,
and MeV fm$^{6}$ for $y_{33}$.}
\begin{ruledtabular}
\begin{tabular}{lddddd}
 		& \text{KIDS0} 		& \text{KIDS-A} 	& \text{KIDS-B} 	& \text{KIDS-C} & \text{KIDS-D} \\ \hline
$t_0$ 	&  -1772.044  & -1855.377    &-1772.044	&  -1688.710	&  -1605.377	\\
$y_0$ 	&     -127.524	&     2302.760	&     2057.283	&      751.561	&     543.801	\\
$t_1$ 	&      300.700	&      300.737	&      296.306	&      295.097	&     293.565	\\
$t_2$ 	&     -176.137	&     -182.308	&    -176.616 	&     -169.851	&    -163.281	\\
$t_{31}$	&   12216.730	&   14058.746	&   12216.730	&   10374.714	&    8532.698	\\
$y_{31}$& -11969.990&-76808.423   &-70716.593	& -33487.909	& -29128.971	\\
$t_{32}$ &      564.638	&   -1020.642	&      620.997	&    2208.265	&    3802.565	\\
$y_{32}$&   29153.150	&  127228.272& 118923.207	&   64791.066	&   59087.582	\\
$y_{33}$& -22955.280& -75362.002& -70290.560 & -46589.511 & -41756.516 \\
$W_{0}$&  116.088& 98.683& 98.929 & 101.306 & 101.664 \\
$t^{\prime (\text{n})}_0 $&   -240.883 & -237.198 &   -238.688	&    -241.695	&  -244.660 \\
$t^{\prime (\text{p})}_0 $&   -262.168 &    -254.737	&    -256.790	&    -260.580	&   -264.239 
\end{tabular}
\end{ruledtabular}
\end{table*}

In the second step, the energy density in Eq. (\ref{eq:matter}) is transformed into a local single-particle potential
that is applicable to the Kohn--Sham scheme.
A multiple non-integer power dependence on density in Eq.~(\ref{eq:matter}) 
accounts for the correlations beyond the mean-field approximation in the sense of DFT. 
A gradient of density and spin-orbit interactions, which are null in homogeneous nuclear matter, play a non-vanishing role in finite nuclei.
Two additional parameters are then introduced. 
In the five KIDS models in Table~\ref{tab:model}, 
the two new parameters are fitted to the binding energy and charge radius of $^{40}$Ca, $^{48}$Ca, and $^{208}$Pb.
We note that the parameters determined in the first step for nuclear matter are kept unchanged in fitting the parameters
of the density gradient and spin-orbit interactions.
Since $^{40}$Ca, $^{48}$Ca, and $^{208}$Pb are doubly-closed shell nuclei, 
the pairing does not affect their properties.

In the third step, the pairing force is determined.
We employ a $\delta$-function form for the pairing interaction as
\begin{align}
V_{\text{pair}}^{\text{(n/p)}} = t_0^{\prime \text{(n/p)}} \left( 1 - \frac{\rho}{2 \rho_0} \right) \delta(\vec{r}_1 - \vec{r}_2).
\end{align}
The parameter $t_0^\prime$ is fitted to the three-point formula for the odd-even staggering
centered at the odd-mass system and averaged over the two neighboring nuclei~\cite{PTPS-146-70}, 
  $[\Delta^{(3)}_{\rm n}(N+1,Z)+\Delta^{(3)}_{\rm n}(N-1,Z)]/2$ and
  $[\Delta^{(3)}_{\rm p}(N,Z+1)+\Delta^{(3)}_{\rm p}(N,Z-1)]/2$ for neutrons and protons, respectively, where
\begin{subequations}
\begin{align}
&\Delta^{(3)}_{\text{n}} (N, Z) = \notag\\
& \frac{(-1)^N}{2} \left[ E(N+1, Z) + E(N-1, Z) - 2 E(N, Z) \right],  \\
&\Delta^{(3)}_{\text{p}} (N, Z) = \notag\\
& \frac{(-1)^Z}{2} \left[ E(N, Z+1) + E(N, Z-1) - 2 E(N, Z) \right].
\end{align}
\end{subequations}
As in Ref.~\cite{yam09}, in this work we fit the pairing parameters to the data of $^{156}$Dy
which are 1.17 MeV and 0.98 MeV for neutrons and protons, respectively.
The corresponding 
Skyrme parameters $t_i$, $y_i$ ($= t_i x_i$) and the pairing parameters $t^\prime_0$ 
for the KIDS models are summarized in Table~\ref{tab:parameter}.
The KIDS models are implemented in the HFBTHO code~\cite{sto13} to solve 
the Kohn--Sham--Bogoliubov equation taking the deformation into account. 
The calculations are performed in the $N_{\text{max}}=16$ full spherical oscillator shells. 
To find the global-minimum solution, we vary the deformation parameter 
of the initial configuration in the range of $-0.3 \leq \beta_2 \leq 0.3$. 
The quasiparticle (qp) states are truncated according to the equivalent single-particle energy cutoff at 60 MeV, 
and the qp states with the magnetic quantum number up to $\Omega = 23/2\mbox{--}29/2$ are 
included depending on the initial $\beta_2$ value.

\section{Results and discussion}\label{result}

\begin{figure*}[t]
\begin{center}
\includegraphics[width=0.85\textwidth]{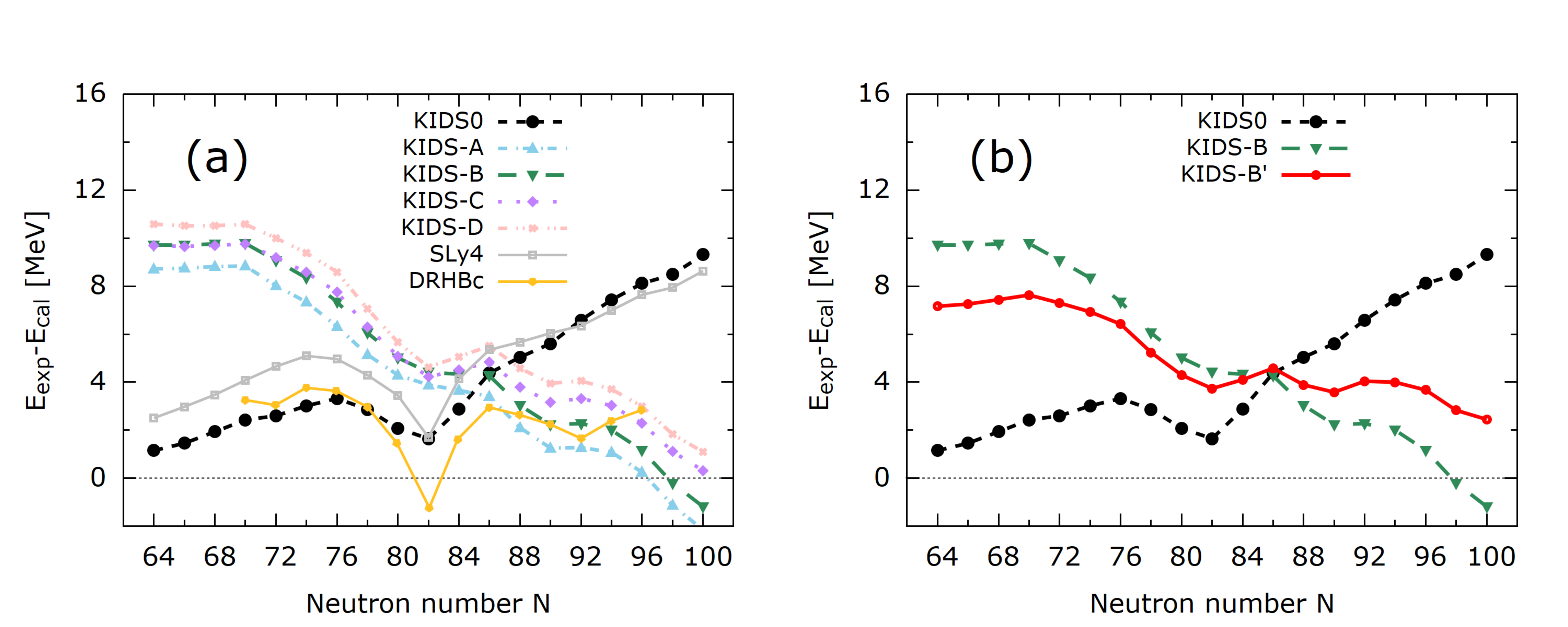}
\end{center}
\caption{\label{fig1}
(a): Difference of the total binding energy between the experiment $E_{\rm exp}$ and the calculation $E_{\rm cal}$. 
(b): Testing the effect of the slope parameter $L$.
The value of $L$ in the KIDS-B model is replaced with that of the KIDS0 model. 
The EoS parameters of the KIDS-B' model are ($K_0$, $J$, $L$, $K_\tau$) = (240, 32, 49, $-420$) given in the units of MeV.
The values of the relativistic DRHBc model are taken from Ref.~\cite{zha20}.}
\end{figure*}

Figure~\ref{fig1}(a) shows the difference of the measured total binding energies ($E_{\rm exp}$) 
from the calculated ones ($E_{\rm cal}$) 
as a function of the neutron number $N$. 
A difference is as significant as 10 MeV. 
The total binding energy of the Nd isotopes is higher than 1000 MeV, 
and the deviation from the experiment is less than 1\% 
even in the worst case. 
The KIDS0 model has a similar isotopic dependence to the SLy4 model. 
A prominent feature is that the slope of the difference is positive for KIDS0 and SLy4 (called Group 1), 
and negative for KIDS-A, B, C, and D (Group 2).
We are going to investigate this behavior. 

For the KIDS0 model, the parameters of the isovector interactions were fitted to the pure neutron matter EoS obtained
by the theoretical calculation~\cite{akm98}.
As for SLy4, the same neutron matter EoS was adopted, and 
the model was further calibrated by the combination of nuclear data and nuclear matter properties.
Each model in Group 2 (KIDS-A, B, C, and D models)
 was optimized by fitting to both nuclear data and neutron star observation.
As one goes from A to D, 
the EoS parameters $K_0$ and $K_\tau$ increase, while $J$ and $L$ decrease. 
On the other hand, Group 1 (KIDS0 and SLy4 models) does not show the complementing behavior of $K_0$ and $K_\tau$, and $J$ and $L$.
It is worthwhile to note that $K_0$ and $J$ values of Group 1 are similar to those of KIDS-A and B, but
$L$ value of Group 1 is smaller than those of KIDS-A and B.
Therefore the difference in $L$ could be a reason for the opposite sign of the slope.
To verify this, we performed a test calculation by changing $L$ value in the KIDS-B model from 58 MeV to 49 MeV that is the value of KIDS0.
We call this test model KIDS-B'.
The results of the KIDS-B' model are compared with those of KIDS0 and KIDS-B in Fig.~\ref{fig1}(b).

The result shown in Fig.~\ref{fig1}(b) demonstrates that the difference in $L$ is one of the sources that contribute to the opposite sign
in the slope of $E_{\rm exp} - E_{\rm cal}$ between Group 1 and Group 2.
It is likely that $E_{\rm exp} - E_{\rm cal}$ could be made flat (zero slope) and close to 0 by adjusting $K_0$, $J$, $L$, and $K_\tau$.
The UNEDF model~\cite{kor10} as well as the relativistic model denoted by DRHBc in Fig.~\ref{fig1}(a)~\cite{zha20}
show a global flat behavior of $E_{\rm exp} - E_{\rm cal}$ with respect to $N$.
Based on the result of KIDS-B', we expect that a global fitting to mass data using a KIDS EDF 
can give a mass table from light to superheavy elements
similar to the UNEDF model in quality. 
The construction of the mass table with a KIDS EDF thus obtained by a global fitting is under consideration.

\begin{figure}[t]
\begin{center}
\includegraphics[width=0.45\textwidth]{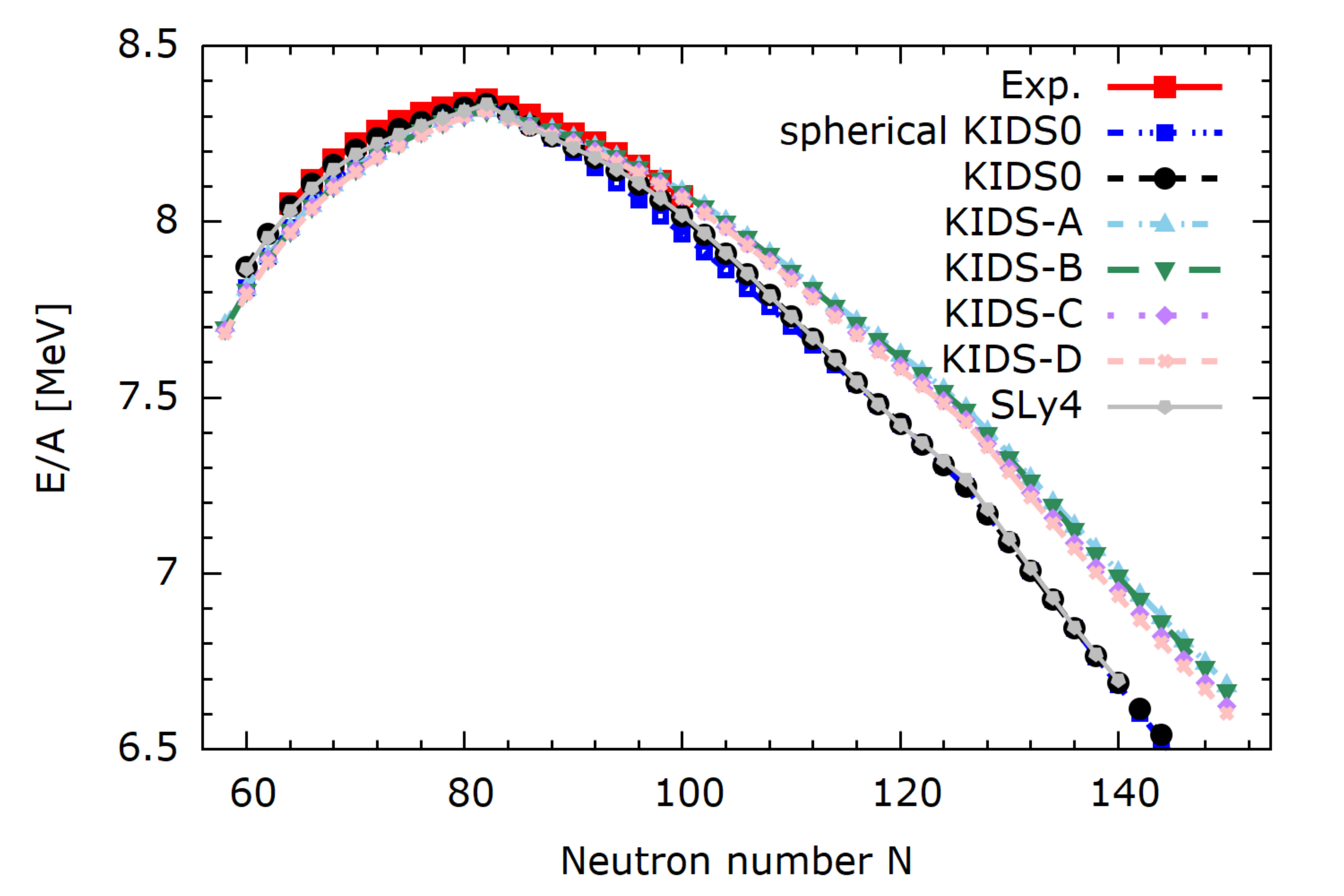}
\end{center}
\caption{Binding energy per nucleon $E/A$ calculated by using the KIDS and SLy4 models 
and comparison with experimental data~\cite{hua17,wan17}. 
Shown are also the results obtained by assuming the spherical symmetry with KIDS0.}
\label{fig3}
\end{figure}

In Fig.~\ref{fig3}, we show the binding energy per nucleon.
According to what is seen in Fig.~\ref{fig1}, 
we obtain a better agreement at $N \leq 82$ with KIDS0 and SLy4,
and better at $N > 82$ with KIDS-A, B, C, and D.
Since the discrepancy between the experiment and calculation is of the order of per mille, 
the agreement to data~\cite{hua17,wan17} is good for all the models.
A notable difference between Group 1 and Group 2 happens at $N > 100$ where experimental data are absent.
The models in Group 2 predict the nuclides that are more bound than those in Group 1.
A difference between the two groups increases as the neutron number evolves.  
At $N=120$ for example, $E/A$ for two groups differs by 0.2 MeV. 
This gives a 36 MeV difference in the total energy, and the difference might be identified in a future measurement.
While the difference between Group 1 and Group 2 is evident, a difference within a group is negligible. 
In Fig.~\ref{fig3}, included are also the results of the KIDS0 assuming the spherical symmetry.
They are denoted by the filled blue squares. 
Except for around $N=100$, it is hard to see the effect of the deformation in KIDS0 on the binding energy.

\begin{figure}[t]
\begin{center}
\includegraphics[width=0.45\textwidth]{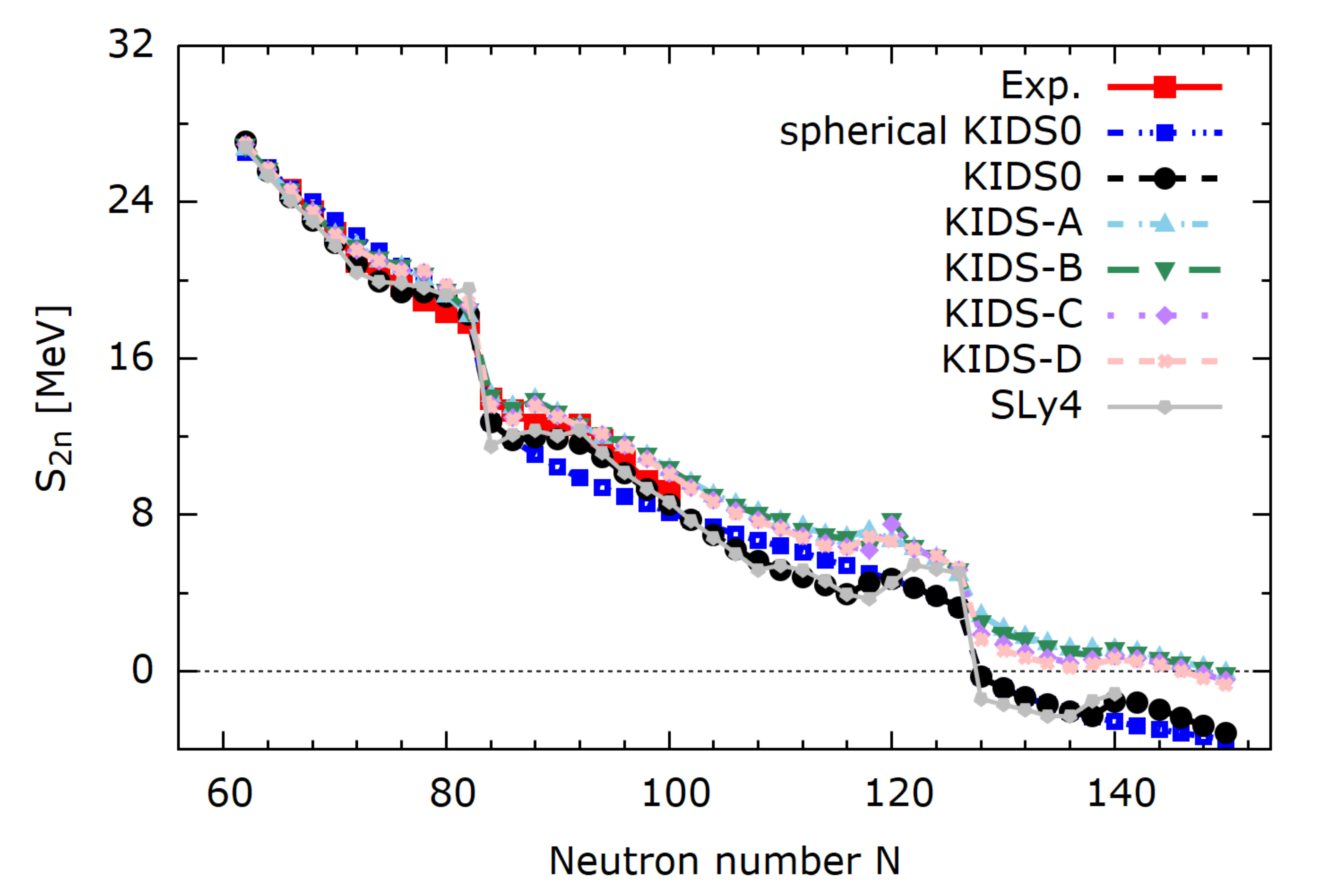}
\end{center}
\caption{Same as Fig.~\ref{fig3} but for two neutron separation energy $S_{2n}$.}
\label{fig4}
\end{figure}

Shown in Fig.~\ref{fig4} is the 
two-neutron separation energy defined by 
\begin{eqnarray}
S_{2n}(N,Z) = E(Z, N) - E(Z, N-2).
\end{eqnarray}
Up to $N=100$, Group 1 and Group 2 behave alike,
and the models deviate from the experimental data within 10\%.
Similarly to $E/A$, a significant difference occurs at $N>100$.
With an increase in the neutron number, 
the Group 1 models show a faster decrease in $S_{2n}$ than Group 2 models do.
As a result, the predicted neutron drip, where $S_{2n}$ becomes negative, is different.
For Group 1, the neutron drip line is located at $N=128$, 
while it extends to $N=148$ for Group 2. 
Around $N=140$, one sees a slight increase in $S_{2n}$. 
This extra binding mechanism is due to the onset of deformation, as seen below.
The KIDS-B' model, which is located in the middle of Group 1 and Group 2 with respect to $E_{\rm exp} - E_{\rm cal}$, 
predicts the drip line at $N=134$. The position is in between the two groups.
Therefore, the measurement of nuclear masses above $N=100$, more preferably $N \geq 110$, 
would give us a novel criterion about the symmetry energy parameters.

The two-neutron separation energies of the Zr isotopes together with the singly closed O, Ca, Sn, and Pb isotopes 
were investigated without deformation in Ref.~\cite{gil20a}. 
An overall trend is in agreement with the experimental data. 
However, a difference from the experimental data is not negligible around $^{100\mbox{--}110}$Zr, 
where the nuclear deformation is suggested to occur 
based on the low $2^+$ energies~\cite{sum11} and the $\beta$-decay half-lives~\cite{nis11,yos13,yos17}.
The nuclear deformation could be the origin of the discrepancy. 
We thus investigate the deformation effect on $S_{2n}$ here. 
The results obtained by assuming the spherical symmetry 
are denoted by the filled blue squares in Fig.~\ref{fig4}.
They show a monotonic behavior as a function of the neutron number. 
One sees that a difference from the experimental data is apparent off the magic number of 82. 
It is clearly shown that taking into account the deformation improves the error of the calculation 
assuming the spherical shape.

\begin{figure}[t]
\begin{center}
\includegraphics[width=0.45\textwidth]{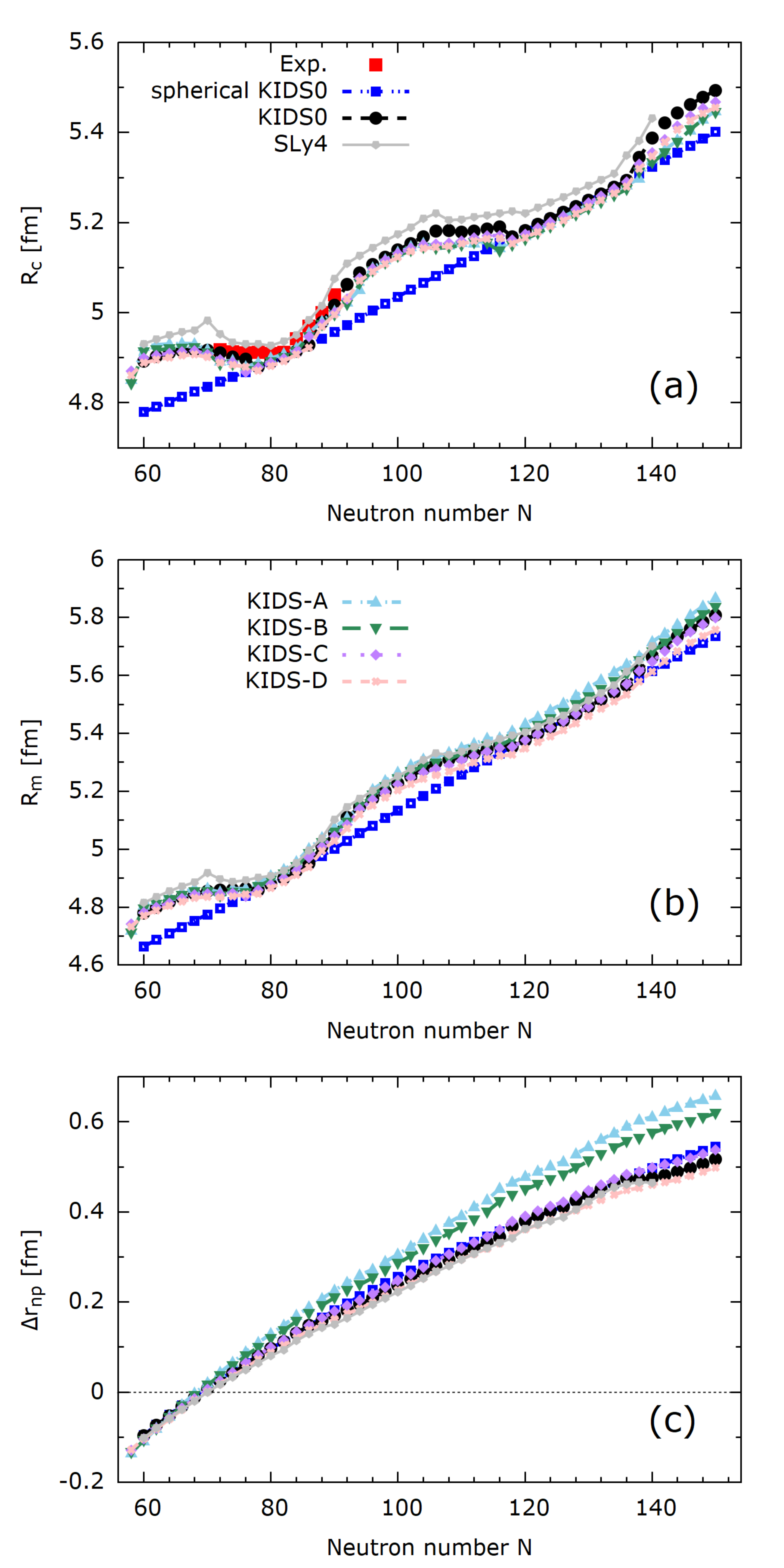}
\end{center}
\caption{\label{fig5} Same as Fig.~\ref{fig3} but for (a) charge radii $R_c$, (b) matter radii $R_{\rm rms,m}$, and 
(c) the neutron-skin thickness $\Delta r_{\text{np}}$.}
\end{figure}

Next, we discuss the radii. 
The top panel of Fig.~\ref{fig5} shows the charge radii calculated 
and compares with the experimental data.
The charge radius $R_c$ is evaluated
from the point-proton radius $R_{\rm rms,p}$ here as 
\begin{align}
R_c=\sqrt{R^2_{\text{rms},\text{p}} + 0.64 \,\text{fm}^2}.
\end{align}

There is no clear distinction between Group 1 and Group 2
over the whole range of the neutron number.
This is at variance with the finding for the binding energy, in which the two groups show 
a considerable discrepancy for $N>100$.
The SLy4 model predicts the radii of the isotopes constantly larger than the KIDS models do,
 and the KIDS models show almost identical results. 
A difference from the experiment is at the level of 0.5\% or less. 
Taking a close look at the figure, 
one sees that the KIDS0 model predicts a larger radius around $N=110$ and $N>140$. 
This can be attributed to the finding that the KIDS0 model predicts a stronger deformation than 
the other KIDS models do, as seen below. 
The results obtained by assuming the spherical symmetry (denoted by the filled blue squares) 
show a considerable difference from those allowing the deformation.
The results of the symmetry restricted and unrestricted calculations 
amount to 0.2 fm difference around $N=100$. 
If the charge radius could be measured with an accuracy of 0.1 fm or less, it could provide evidence for
the deformation of the Nd isotopes. 
In the region where the experimental data are available, 
it is easily seen that the deformation is indispensable to consider 
for a quantitative description of the charge radii.

One can see a different model dependence for 
the matter radii $R_{\text{rms,m}}$ and the neutron-skin thickness $\Delta r_{\text{np}}$, as depicted 
in Figs.~\ref{fig5}(b) and \ref{fig5}(c). 
The calculation employing KIDS-A gives the largest $R_{\rm rms, m}$ and $\Delta r_{\rm np}$, 
and they decrease in the order of KIDS-A, B, C, and D.  
Since the parameters of EoS vary in these models, 
it is not easy to disentangle the major source giving this trend. 
Although KIDS-A has a similar $K_0$ value to SLy4 ($\sim 230$ MeV), 
KIDS-A produces appreciably larger $R_{\text{rms,m}}$ and $\Delta r_{\text{np}}$ than SLy4 does. 
Thus, $K_0$ is unlikely to govern the model dependence of the matter radii. 
Since the values of $K_\tau$ are the same between KIDS-A and B, and
between KIDS-C and D, 
$K_\tau$ is also unlikely to explain the model dependence. 
The combination of $J$ and $L$ would be responsible, 
and a further systematic study is needed.

\begin{figure}[t]
\begin{center}
\includegraphics[width=0.45\textwidth]{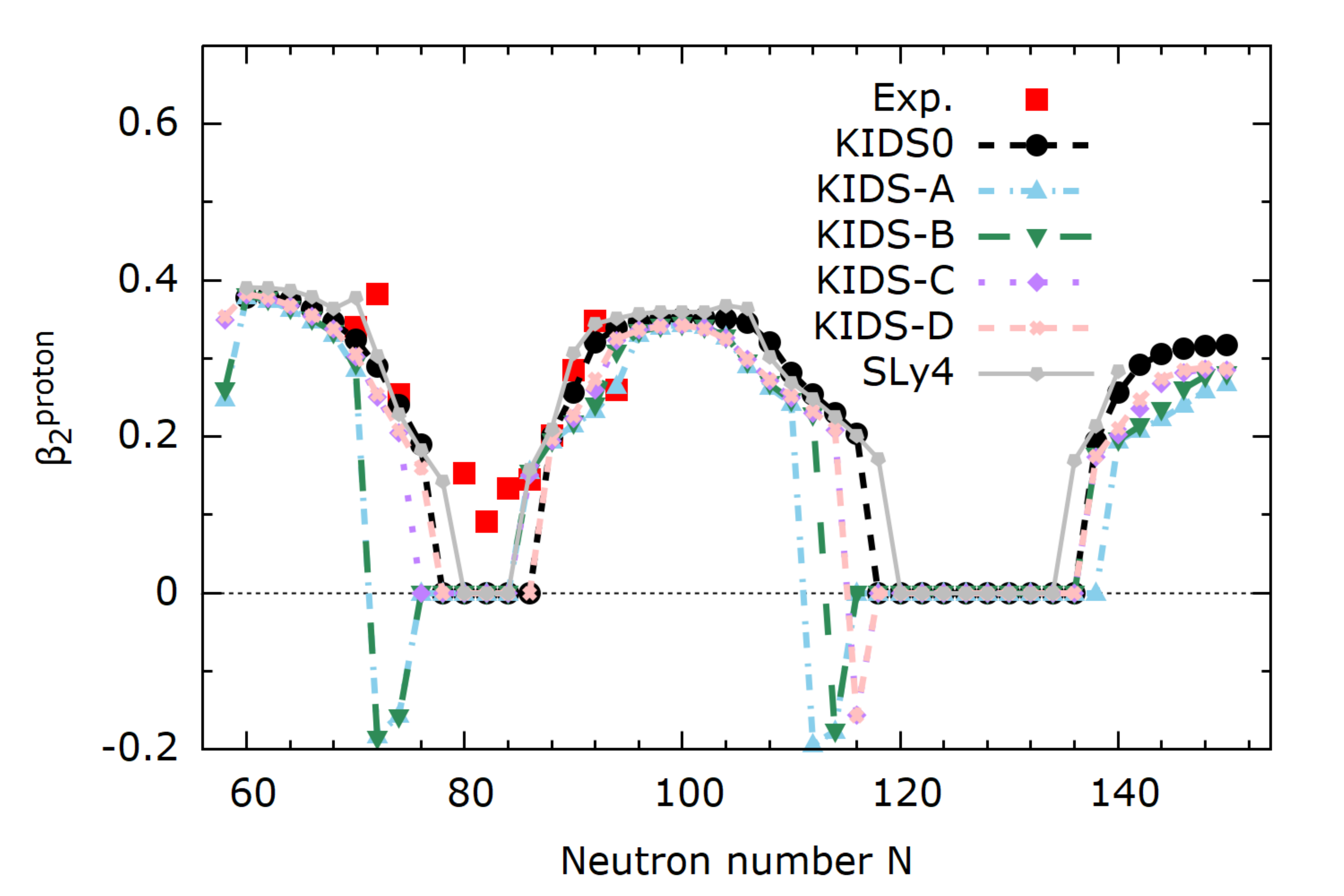}
\end{center}
\caption{\label{fig6} Same as Fig.~\ref{fig3} but for quadrupole deformation $\beta_{2,{\rm p}}$.
}
\end{figure}

Figure~\ref{fig6} shows 
the quadrupole deformation parameter $\beta_{2,{\rm p}}$ defined by
\begin{align}
\beta_{2,\text{p}}=\dfrac{\sqrt{5\pi}Q_{2,\text{p}}}{3Z R^2_{\text{rms},\text{p}}}
\end{align}
for protons, where $Q_{2,\text{p}}$ stands for the proton quadrupole moment. 
The agreement to experiment~\cite{ram01} is reasonable. 
Similarly to the charge radii, 
the dependence of the quadrupole moment on the model is weak. 
As discussed above, 
one sees that the strongly deformed isotopes obtained by the SLy4 and KIDS0 models have larger charge radii. 
We observe the onset of deformation around $N=138$ near the drip line, 
which extends the drip line toward large $N$ thanks to the increased binding.

It is also noted that around the spherical magic number of $N=82$, 
a difference between the calculated and measured deformation is apparent. 
As discussed in Ref.~\cite{yos11b}, 
the discrepancy is due to the lack of a dynamical deformation effect in the present calculation. 
Extension of the present study to include the dynamical effect based on the HFBTHO code 
as in Refs.~\cite{los10,sto11,ois15} is thus an interesting future study.
Furthermore, 
an abrupt change of the sign happens at the boundary between the spherical and non-spherical shapes. 
As the neutron number approaches the boundary of the shape transition,
the binding energies of spherical, prolate, and oblate shapes
become close to each other.
The difference is only about $10^{-5}$ of the total binding energy. 
Therefore, 
a unique determination of the ground state deformation is
obscure at the boundary of the shape transition. 

\begin{figure}[t]
\begin{center}
\includegraphics[width=0.45\textwidth]{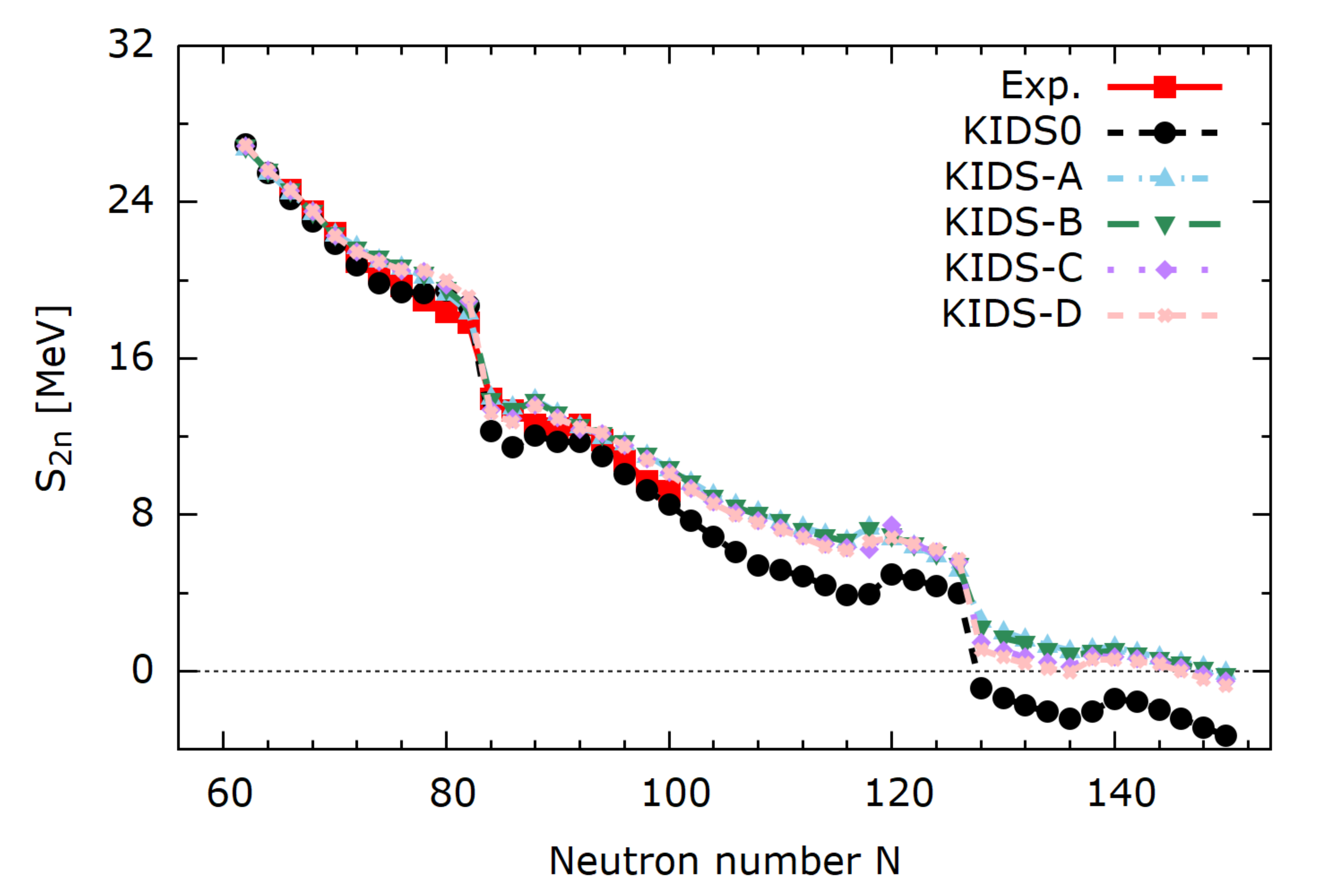}
\end{center}
\caption{\label{fig7} Two neutron separation energy $S_{2n}$ with the pairing force fitted to $^{120}$Sn.}
\end{figure}

Let us discuss the pairing effects briefly because 
the location of the drip line is sensitive to the pairing~\cite{afa15}, 
and the onset of deformation is a consequence of the competing 
particle--hole and particle--particle correlations.
To see the uncertainty due to the pairing force adopted, 
we consider the pairing parameters fitted to the three-point value of
$^{120}$Sn, 1.32 MeV, for neutrons.
The resultant pairing strengths are $t_0^{\prime (\text{n})}=-222.3, -232.4, -232.8, -232.2$, and $-233.5  $ MeV fm$^3$ 
for KIDS0, KIDS-A, B, C, and D, respectively.
These are not very different from the strength determined by using $^{156}$Dy.
For protons, we use the same pairing strength fitted to $^{156}$Dy. 
We thus find that the ground-state properties 
discussed above are essentially the same between the results obtained by employing 
the different pairing strengths;
the neutron number dependence of the binding energy is classified 
into the same two groups, one with KIDS0 and SLy4, and the other with KIDS-A, B, C, and D, and the
calculated size ($R_c$) and shape ($\beta_{2,{\rm p}}$) also agree with the measurements 
with the same quality. 
As an example of the results, we show in Fig.~\ref{fig7} 
the two neutron separation energy $S_{2n}$ calculated with the pairing force that is fitted to neutrons of $^{120}$Sn.
The predicted neutron drip line is located at $N=128$ and 148 in KIDS0 and Group 2, respectively.
For the KIDS0 model, the $S_{2n}$ value at $N=128$ is changed from $-0.30$ MeV to $-0.90$ MeV 
by varying the pairing strength. 
Therefore, if we strengthen the pairing interaction more, the drip line would be extended.

\section{Summary}\label{summary}

The KIDS functional 
has been applied to the Nd isotopes with the neutron numbers ranging from 60 to 160 
in the framework of the nuclear EDF method, 
where the Kohn--Sham--Bogoliubov equation is solved taking the deformation into account. 
We demonstrated that the KIDS functionals 
constructed to satisfy both neutron-matter EoS 
or neutron star observation and selected nuclear data
describe well the ground-state properties 
with respect to the existing data and predictions.
We found that considering the nuclear deformation improves the description of the binding energies and radii. 
In very neutron-rich isotopes with $N>100$, 
the KIDS functionals predict differences in the mass, $S_{2n}$, and $\Delta r_{\text{np}}$ 
though all the models satisfy the neutron star observation.  
A global fitting of the EDF will offer a microscopic mass table for the whole region of the nuclear chart. 
A systematic calculation of the $\beta$- and $\gamma$-strength functions employing the thus obtained EDF is a 
future direction of our study.

\section*{Acknowledgments}
This work was supported by 
the NRF grants funded by the Korean government 
(No. 2018R1A5A1025563 and No. 2020R1F1A1052495), 
the JSPS KAKENHI (Grants No. JP19K03824, No. JP19K03872, No. JP19KK0343, and No. JP20K03964), 
and the JSPS/NRF/NSFC A3 Foresight Program ``Nuclear Physics in the 21st Century.''

\bibliography{KIDS_deformation}

\end{document}